# BSTree: an Incremental Indexing Structure for Similarity Search and Real Time Monitoring of Data Streams

Abdelwaheb Ferchichi, Mohamed Salah Gouider

**Higher Institute of Management, Department of Computer Science, Tunis, Tunisia**
`ferchichiabdelwaheb@yahoo.fr`
`ms.gouider@isg.rnu.tn`

**Abstract.** In this work, a new indexing technique of data streams called BSTree is proposed. This technique uses the method of data discretization, SAX [4], to reduce online the dimensionality of data streams. It draws on Btree to build the index and finally uses an LRV (least Recently visited) pruning technique to rid the index structure from data whose last visit time exceeds a threshold value and thus minimizes response time for similarity search queries.

## 1   Introduction

In this paper, a new data stream indexing structure called BSTree is proposed. It inspired by the recent works of Camerra et al. [2], as well as those of Rudolf Bayer on the Btree. The proposed technique uses the symbolic discretization method, SAX [4], to reduce the dimensionality of the data streams, the sliding windows to perform an online processing of incoming streams. Several changes were also made to the Btree structure to adopt the characteristics of the data streams.

## 2   Contribution

### 2.1   BSTree : Balanced Stream Tree

The new approach is based on:

   **a)  SAX for Summarization and Discretization of the Data Stream.**
   SAX representation is detailed in [4], in this sub-section, we would merely introduce the motivations of our choice, which can be summarized as follows:

- The ease of implementation of SAX, unlike other techniques such as DFT and DWT.

- The existence of a large number of algorithms and data structures that allow efficient handling of symbolic representations.
- The symbolic nature of the SAX representation allows the use of the lexicographical order for sorting the data in the BSTree structure.
- The distance measurement MinDist defined by the authors of SAX between two symbolic strings is very close ($\leq$) to the distance between the original data streams.

**b) A windowing system to extract the features of the data stream.**

To browse the continuous data stream for extracting subsets of elements that will be discretized with "SAX" before being inserted into the BSTree index structure, we use a Sliding Window. Whenever w elements are observed and included in the Sliding Window, a new symbol "SAX" is generated and inserted into the BSTree structure.

**c) The indexing structure BSTree.**

A BSTree of order m has the following properties:

- The root either is a leaf node or has at least two non-empty sub-trees and at most m non-empty sub-trees.
- Each internal node has at least $\lceil m/2 \rceil$ non-empty sub-trees and at most m non-empty sub-trees ($\lceil m/2 \rceil$ is the smallest integer $> m/2$).
- The number of MBRs in each non-leaf node is one less than the number of non-empty sub-trees of this node.
- Each MBR has a predefined number c of distinct symbols (no duplicates) which are sorted in lexicographical order.

## 2.2 Index Building.

The algorithm Build_Index, in table 1, consists of an insertion procedure to construct the index structure in a single pass and incrementally, and a procedure for removal or more specifically for pruning, to maintain the size of the BSTree structure and respect the constraint on the size of the memory.

The pruning procedure uses the LRV technique (Least Recently Visited). The insertion process continues until the size of the BSTree reaches a maximum height, defined by the user. The BSTree is then pruned by deleting branches whose last visit time exceeds a threshold value tmpTh specified in advance and whose variation may affect the quality of the BSTree index structure after the pruning phase and also construction time of the index.

**Table 1.** Algorithm Build_Index

```
Algorithm  Build_Index (BSTree T,Stream S, c,w, m,
htree, Curr_htree, tmpTh )
// S : The data stream to be indexed
// T : The BSTree index structure
```

```
//  c : capacity of the MBR
// w : size of the sliding window
/* m : Order of BSTree, it is the maximum number of ele-
ments (MBRs) per node */
// htree : The maximum height of the BSTree  structure
/* Curr_htree : height of BSTree when calling  to
  Build_Index algorithm
/* tmpTh : The threshold used in the pruning procedure
"LRV-Pruning" */
Begin
  While Curr_htree  ≤  htree do
    Curr_htree =BSTree_Insert(T,S,c,w,m) ;
  end While
  LRV-Pruning(T, tmpTh);
  Curr_htree = getHauteur(T) ;
  // Recursive call to Build_Index procedure
  Build_Index (T,S,c,w,m,htree,Curr_htree ,tmpTh)
end
```

The Build_Index procedure begins with an iterative call to another procedure, BSTree_Insert, used to insert new SAX symbols created until reaching the maximum height of the tree defined by the user. Then we move to a pruning phase of the BSTree index structure based on a threshold value tmpTh of the last visit time per node (call to the procedure LRV-Pruning). Finally we go back to the construction phase by recursively calling the Build_Index procedure which ensures an incremental construction of the BSTree index structure..

**a) The Insertion Procedure BSTree_Insert**

In the BSTree_Insert procedure, a sliding window is used to retrieve the values of the data stream . The SAX procedure uses this window to build the SAX symbol . If the MBR covering this new symbol exists in the index structure, then the SAX symbol will be inserted into the MBR using the procedure MBR_insert. Otherwise, the MBR will be searched in a file that contains all possible combinations of the alphabet according to the size of the MBR and the created symbol is then inserted into the MBR before the latter is inserted into the BSTree index structure using the procedure Index_insert .

*(i) The Procedure MBR_Insert*

The MBR_insert procedure operates the ascending lexicographic order of symbols in an MBR. The procedure browses the MBR until finding the first element lexicographically greater than the newly built symbol. The new symbol is then inserted into this position by performing a translation of the elements of the MBR that are higher than this symbol.

*(ii) The procedure Index_Insert*

The procedure Index_insert is the same as the standard insertion procedure used in B-tree, except that the comparison between elements when traversing the exploits the lexicographic order of the symbols.

*(iii) The procedure SAX*

The SAX procedure is detailed in [4].

**b) The pruning procedure LRV-Pruning**

In the index structure BSTree, a timestamp Ts is associated with each element in a node N. This timestamp is updated at each visit of the corresponding element during a browse of the tree triggered by a query Q.

During the pruning phase, we use a depth-first search algorithm which browses the tree from left to right and with backtracking. The intuition behind this choice is that the traversing of the BSTree index structure is done from top to bottom (from the root to the leaves) in the tree and from left to right within each node, so for two successive elements of the BSTree structure, two cases may arise during the pruning process:

- Either the timestamp $Ts_i$ of the current node is greater than the pruning threshold tmpTh, in this case we must continue the traversing of BSTree structure;
- Either the timestamp $Ts_i$ of the current node is less than the pruning threshold tmpTh, in this case also two sub-cases are considered:
  - The timestamp $Ts_i$ of the current node is less than the timestamp $Ts_{i+1}$ of the next node, so there is a possibility to reach nodes with timestamps equal or superior to the pruning threshold tmpTh and we should, therefore pursue the traversing of the BSTree structure and thus maintain this "bridge" of nodes that can lead us to the other nodes that do not obey to the pruning condition;
  - The timestamp $Ts_i$ of the current node is greater than the timestamp $Ts_{i+1}$ of the next node, in this case we must perform a backtracking then prune this branch before continuing the traversal of the tree.

Before detailing the pruning algorithm, it is important to note that:

- During the insertion Phase "BSTree_Insert", any new insertion is performed at the leaf nodes, the timestamp Ts of the newly inserted element is initialized to zero, but in special cases where a balancing of the BSTree index structure is needed, the new element can be inserted in a non-leaf node. In this particular case, the timestamp Ts of the newly inserted element is initialized to the maximum of the timestamps of elements of the children node, and this in order to maintain the sort of the elements of each path in the BSTree structure initially based on the values of the timestamps;

- After each pruning phase, all timestamps Ts are set to zero.

The LRV-pruning procedure begins by recovering the root of the tree T. This value of the root of the tree T and the pruning threshold tmpTh are passed as the effective parameters to the DFS procedure used to traverse the tree using the depth-first technique with backtracking.

The DFS procedure is composed of two phases:

- A search phase of the pruning target node,
- A pruning phase then continues the procedure by referring back to the first phase until pruning all the target branches.

The tree obtained after the pruning phase is not balanced. One of the solutions proposed to our BSTree structure to be balanced after the pruning procedure is to insert the unpruned elements in a new structure that will replace the old one, which will be destroyed at the end of the pruning phase.

## 3    Experimental Results and Performance Evaluation

The new indexing technique, BSTree, is compared to Stardust [1].

**Evaluation of the Precision.**

The relationship between the precision and the variation of the radius for range queries is first studied, while setting the size of the basic window TW to 512 and the number of basic windows processed NW to 3600.

Figure 1 shows values of precision for BSTree which are higher than those of Stardust for radii ranging from 0.1 to 1, but still more many other better precision values for BSTree after the pruning stage, this is due to the elimination of unnecessary nodes during the pruning phase which will increase the quality of search operations.

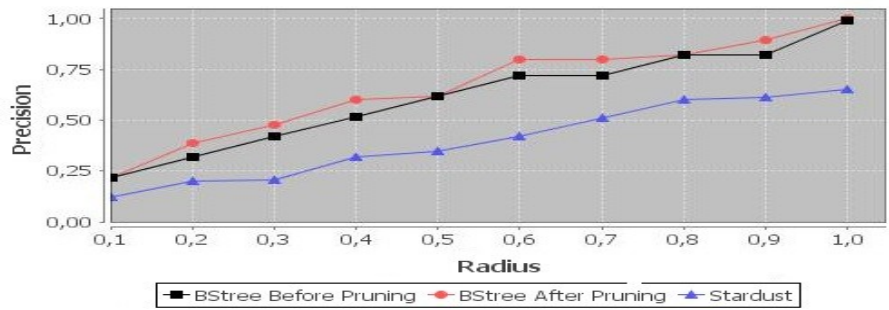

**Fig. 1.** Precision of BSTree before and after pruning phase VS Stardust by varying the radius of the queries for the "packet.dat"[3] dataset

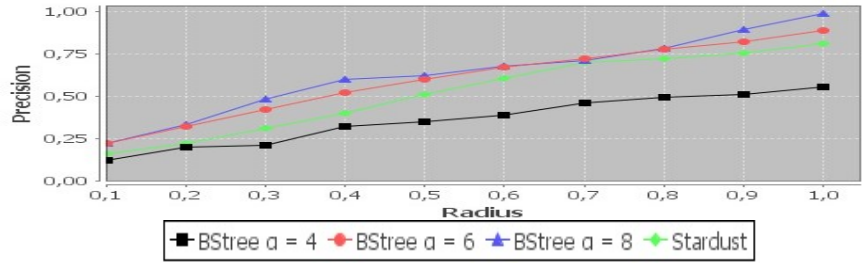

**Fig. 2.** BSTree precision for sizes of the alphabet SAX α = 4, α = 6 and α = 8 VS Stardust by varying the radius of the queries for the synthetic dataset

In Figure 2, the relationship between the precision and the size of the alphabet used in the discretization phase (SAX) was studied. We note that Stardust performs more than BSTree for an alphabet of size α = 4. For sizes of the alphabet greater than 4 (α = 6 and α = 8), the precision of BSTree is better than Stardust and can reach values close to 1. This can be explained as follows: If the size of the alphabet increases, the quality of SAX transformation becomes better (the number of windows of data streams that correspond to the same symbol SAX decreases) and vice versa, which will affect the quality of the results produced by BSTree.

**Evaluation of the Recall.**

Closely to the precision, we also note that BSTree returns values of recall superior to those of Stardust.

## 4    Conclusion

In this paper, we have proposed a new structure of incremental indexing for similarity search and real-time monitoring of data streams that we have called BSTree. It is based on the technique of symbolic dimensionality reduction SAX and the famous Btree indexing structure.

## References

1. Bulut, A., Singh, A.K.: A unified framework for monitoring data streams in real time. Data Engineering, 2005. ICDE 2005. Proceedings. 21st International Conference on. pp. 44–55 (2005).
2. Camerra, A. et al.: iSAX 2.0: Indexing and mining one billion time series. Data Mining (ICDM), 2010 IEEE 10th International Conference on. pp. 58–67 (2010).
3. Keogh, E. Zhu, Q. Hu,B. Hao, Y.Xi, X.Wei, L and Ratanamahatana, C. A. : The UCR Time Series Classification/Clustering Homepage: www.cs.ucr.edu/~eamonn/time_series_data/.( 201.
4. Lin, J. et al.: Experiencing SAX: a novel symbolic representation of time series. Data Mining and Knowledge Discovery. 15, 2, 107–144 (2007).